\documentclass[aps, pre, twocolumn, superscriptaddress]{revtex4-1}

\usepackage{amsmath}
\usepackage{amssymb}
\usepackage{xspace}
\usepackage{color}
\usepackage{graphicx}

\usepackage{ifpdf}
\ifpdf
\fi

\newcommand{\eq}[1]{(\ref{#1})}
\newcommand{\Eq}[1]{Eq.~(\ref{#1})}
\newcommand{\Eqs}[1]{Eqs.~(\ref{#1})}
\newcommand{\Fig}[1]{Fig.~\ref{#1}}

\newcommand{\Ref}[1]{Ref.~\cite{#1}}
\newcommand{\Refs}[1]{Refs.~\cite{#1}}

\newcommand{\eg}{{e.g.,\/}\xspace}
\newcommand{\ie}{{i.e.,\/}\xspace}

\newcommand{\pd}{\partial}

\newcommand{\mc}[1]{\mathcal{#1}}
\newcommand{\mcc}[1]{\mathfrak{#1}}
\renewcommand{\vec}[1]{{\boldsymbol{\rm #1}}}
\newcommand{\favr}[1]{\langle #1 \rangle}

\newcommand{\moyalsin}[1]{\{\!\{#1\}\!\}}
\newcommand{\moyalcos}[1]{[[#1]]}

\begin{document}

\title{{Wave kinetics of drift-wave turbulence and zonal flows beyond the ray approximation}}
\author{Hongxuan Zhu}
\affiliation{Department of Astrophysical Sciences, Princeton University, Princeton, New Jersey 08544, USA}
\affiliation{Princeton Plasma Physics Laboratory, Princeton, New Jersey 08543, USA}

\author{Yao Zhou}
\affiliation{Princeton Plasma Physics Laboratory, Princeton, New Jersey 08543, USA}

\author{D.~E. Ruiz}
\affiliation{Sandia National Laboratories, P.O. Box 5800, Albuquerque, New Mexico 87185, USA}

\author{I.~Y. Dodin}
\affiliation{Department of Astrophysical Sciences, Princeton University, Princeton, New Jersey 08544, USA}
\affiliation{Princeton Plasma Physics Laboratory, Princeton, New Jersey 08543, USA}

\begin{abstract}
Inhomogeneous drift-wave turbulence can be modeled as an effective plasma where drift waves act as quantumlike particles and the zonal-flow velocity serves as a collective field through which they interact. This effective plasma can be described by a Wigner--Moyal equation (WME), which generalizes the quasilinear wave-kinetic equation (WKE) to the full-wave regime, i.e., resolves the wavelength scale. Unlike waves governed by manifestly quantumlike equations, whose WMEs can be borrowed from quantum mechanics and are commonly known, drift waves have Hamiltonians very different from those of conventional quantum particles. This causes unusual phase-space dynamics that is typically not captured by the WKE. We demonstrate how to correctly model this dynamics with the WME instead. Specifically, we report full-wave phase-space simulations of the zonal-flow formation (zonostrophic instability), deterioration (tertiary instability), and the so-called predator--prey oscillations. We also show how the WME facilitates analysis of these phenomena, namely, (i) we show that full-wave effects critically affect the zonostrophic instability, particularly, its nonlinear stage and saturation; (ii) we derive the tertiary-instability growth rate; and (iii) we demonstrate that, with full-wave effects retained, the predator--prey oscillations do not require zonal-flow collisional damping, contrary to previous studies. We also show how the famous Rayleigh--Kuo criterion, which has been missing in wave-kinetic theories of drift-wave turbulence, emerges from the WME.
\end{abstract}

\maketitle

\section{Introduction} 
Drift waves (DWs) in plasma physics and mathematically similar Rossby waves in geophysics can spontaneously generate coherent nonlinear structures in the form of banded shear flows, which are commonly known as zonal flows (ZFs). Interactions between ZFs and DW turbulence are a fundamental problem that has been actively studied for decades, particularly due to its importance for turbulent transport in magnetic-fusion devices \cite{foot:review}. A common model for studying these interactions is the wave-kinetic equation (WKE) \cite{ref:diamond05, ref:smolyakov99, ref:smolyakov00a, ref:smolyakov00b, ref:malkov01b, ref:kim02, ref:kaw02, ref:trines05, ref:miki12, ref:diamond94, ref:kim03, ref:singh14}, which relies on the geometrical-optics (GO) approximation, \ie the characteristic DW wavelength is assumed small compared to the ZF scales. However, this assumption is not always justified \cite{ref:fujisawa04, ref:gupta06, ref:hillesheim16, ref:st-onge17}, and essential physics is lost in the GO limit \cite{ref:parker16, my:zonal}. (Additional evidence is also presented below.) This stimulated formulations of full-wave statistical theories, which remain manageable within the quasilinear approximation, i.e., when eddy-eddy interactions are ignored. A particularly notable example is the second-order cumulant expansion (CE2), which has been used in both geophysics and plasma physics \cite{ref:parker13, ref:parker14, ref:srinivasan12, ref:constantinou16}. However, the CE2 is formulated in terms of the two-point correlation function, so it is not an obvious generalization of the WKE, which describes the DW dynamics in the ray phase space. Thus, an alternative theory is needed to unify the WKE and the full-wave approach to inhomogeneous turbulence.

Recently, it was noticed \cite{my:zonal} that DWs can be viewed as effective \textit{quantum} particles for which the ZF velocity serves as a collective field. Then the DW Wigner function serves as a quasiprobability distribution of DW quanta (``driftons'') in phase space. It fully determines the ZF dynamics and satisfies a kinetic equation of the Wigner--Moyal (WM) type \cite{ref:moyal49, ref:wigner32}. This leads to a complete model of DW turbulence in the same quasilinear approximation that underlies the CE2 (and was also extended recently beyond the quasilinear approximation \cite{ref:ruiz18}). However, unlike the CE2, the WM model describes the dynamics in phase space; thus, it leverages the existing Hamiltonian formalism and provides a connection with the WKE, which is subsumed as the GO limit. Previous applications of this full-wave phase-space approach to classical turbulence have been restricted to manifestly quantumlike systems such as those governed by the nonlinear Schr\"odinger equation \cite{ref:hall02, ref:onorato03, ref:semenov08, ref:eliasson10, ref:hansson12, ref:hansson13, ref:picozzi14} (\eg optical turbulence in Kerr media) and the Klein-Gordon equation \cite{ref:santos05, ref:santos07}. In those cases, the WM equations are basically borrowed from quantum mechanics and reduce to the commonly known WKEs in the GO limit. In contrast, driftons have Hamiltonians very different from those of conventional quantum particles and are also subject to dissipation even in a collisionless plasma. This causes unusual phase-space dynamics and makes the GO approximation a subtle matter. In particular, it was found that the GO limit of the WM equation for driftons is not \textit{quite} the traditional WKE (tWKE) but includes corrections that reinstate the conservation of the DW--ZF total enstrophy. Applications of this ``improved'' wave-kinetic equation (iWKE) \cite{foot:iwke} were contemplated in Refs.~\cite{ref:parker16, my:zonal}, but the utility of full-wave WM modeling of DW turbulence has not been explored yet.

Here we report the full-wave phase-space modeling of inhomogeneous DW turbulence as an effective quantumlike drifton plasma. The general mathematical formulation of the WM equation is taken from Ref.~\cite{my:zonal} and the plasma is assumed collisionless for simplicity. We simulate the ZF formation [zonostrophic instability (ZI) \cite{ref:smolyakov00a, ref:smolyakov00b, ref:srinivasan12, ref:parker13, ref:parker14}], deterioration [tertiary instability (TI) \cite{ref:kim02, ref:numata07, ref:singh16, ref:st-onge17, ref:rogers00, ref:rogers05}], and the DW--ZF predator--prey-type oscillations \cite{ref:malkov01b, ref:miki12, ref:diamond94, ref:kim03}. We also show how the WM approach facilitates analysis of these effects. Our specific findings are the following. (i) For the linear stage of the ZI, when the tWKE dynamics is simulation-box dependent, the WM model predicts physical rates that account for full-wave effects and agree with the CE2. The accuracy of the corresponding iWKE predictions is, in general, only qualitative. (ii) The iWKE predicts three types of drifton phase-space trajectories. Our analysis of these trajectories shows that adequate modeling of the ZI nonlinear stage and saturation requires accounting for full-wave effects, which is impossible within both the tWKE and the iWKE. (iii) When full-wave effects are retained, predator--prey oscillations do not require ZF collisional damping, contrary to  previous studies. Moreover, we find that these oscillations occur in our simulations only outside the validity domain of their tWKE-based existing theory. (iv) The TI cannot be described by the tWKE or the iWKE in principle, but it is captured by the WM analysis. We calculate the TI growth rate and compare our results with simulations. (v) The famous Rayleigh--Kuo criterion \cite{ref:kuo49}, which is known from geophysics yet has been missing in tWKE-based theories, emerges after full-wave corrections are reinstated. 

Overall, our work corrects and extends previous efforts in phase-space studies of inhomogeneous DW turbulence such as in Ref.~\cite{ref:trines05}. Hence, it serves as a step toward revising basic physics of DW turbulence (and potentially, its impact on turbulent transport) from a different perspective. The specific findings reported here are only intended to illustrate the utility of the general WM formulation in application to inhomogeneous DW turbulence. Likewise, the specific turbulence model used below is just an example chosen for its simplicity and relevance to the existing tWKE and CE2 models. Wigner--Moyal studies of DW turbulence within more realistic models is something that this work seeks to stimulate in the future.   

\section{Basic equations}
The plasma model adopted in this paper is as follows. We assume cold ions, electrons with temperature $T_e$, and a uniform magnetic field $\vec{B}_0 = B_0 \hat{\vec{z}}$, where $\hat{\vec{z}}$ is a unit vector along the $z$ axis. The equilibrium density gradient $\nabla n_0$ is in the $y$ direction. The electrostatic potential $\varphi$ is described by the generalized Hasegawa--Mima equation (gHME) \cite{ref:krommes00, ref:smolyakov00a, ref:smolyakov00b, ref:hammett93}
\begin{gather}\notag
\pd_t w + (\hat{\vec{z}}\times\nabla\varphi) \cdot \nabla w + \beta\,\pd_x\varphi = 0,
\quad
w = (\nabla^2-\hat{a})\varphi
\end{gather}
for the generalized vorticity $w(t, \vec{x})$ on the $\vec{x} \equiv (x, y)$ plane transverse to $\vec{B}_0$. Here, time is measured in units $\Omega_i^{-1}$, where $\Omega_i$ is the ion gyrofrequency; length is measured in units $\rho_s\doteq c_s/\Omega_i$ ($\doteq$ denotes definitions), where $c_s$ is the ion sound speed; $\varphi$ is measured in units $T_{e}/|e|$, where $e$ is the electron charge; also, $\beta$ is proportional to $\pd_y n_0$ and is treated as a positive constant. The operator $\hat{a}$ models the electron response to $\varphi$ such that $\hat{a}=1$ for DWs and $\hat{a}=0$ for ZFs \cite{ref:hammett93}. External forcing and dissipation are not included because they are not directly relevant to the effects discussed below. ({If the stochastic forcing were retained, ergodicity in the $x$ direction would have to be assumed, like in the CE2 \cite{ref:srinivasan12}.}) For any given $f$, we introduce its zonal average $\favr{f}\doteq\int fdx/L_x$ ($L_x$ is the system length in the $x$ direction) and fluctuations $\smash{\widetilde{f} \doteq f - \favr{f}}$. ZFs are described by the average velocity $U(t, y)\doteq-\favr{\varphi'}$. (Primes denote derivatives with respect to $y$.) {Assuming the quasilinear approximation}, DWs are governed by  $i\pd_t \widetilde{w} = \hat{H}\widetilde{w}$, where $\hat{H}$ serves as the drifton Hamiltonian \cite{my:zonal}. We also introduce the zonal-averaged Wigner function $W(t, y, \vec{p})\doteq \favr{\int e^{-i\vec{p}\cdot\vec{s}}\, \widetilde{w}(t, \vec{x}+\vec{s}/2) \widetilde{w}(t, \vec{x}-\vec{s}/2)\,d^2s}$. Then, the WM formulation is \cite{my:zonal}
\begin{gather}
\pd_t W = \moyalsin{\mc{H}, W} + \moyalcos{\Gamma, W}, \label{eq:WM_W}
\\
\textstyle
\pd_t U = \pd_y \int p_{D}^{-2}\star p_xp_yW\star\,p_{D}^{-2}\,d^2p/(2\pi)^2.\label{eq:WM_U}
\end{gather}
Here, $\mc{H}$ and $\Gamma$ are the Weyl symbols of the Hermitian and anti-Hermitian parts of $\hat{H}$:
\begin{gather}
\mc{H}=-\frac{\beta p_x}{p_D^2} + p_x U + \frac{1}{2}\,[[U'',\frac{p_x}{p_D^2}]], 
\quad
\Gamma=\frac{1}{2}\,\moyalsin{U'',\frac{p_x}{p_D^2}},
\notag
\end{gather}
where $p_D^2\doteq1+p_x^2+p_y^2$. Also, $\star$ is the Moyal star, $\smash{A \star B \doteq Ae^{i\hat{\mc{L}}/2}B}$, where $\smash{\hat{\mc{L}} \doteq \overleftarrow{\pd_{\vec{x}}} \cdot \overrightarrow{\pd_{\vec{p}}} - \overleftarrow{\pd_{\vec{p}}} \cdot \overrightarrow{\pd_{\vec{x}}}}$ and the arrows indicate the directions in which the derivatives act. For example, $A\hat{\mc{L}}B$ is the canonical Poisson bracket, $\{A,B\}$. Also, $\moyalsin{A,B} \doteq 2A\sin(\hat{\mc{L}}/2)B$ and $\moyalcos{A, B}\doteq 2A\cos(\hat{\mc{L}}/2)B$. We solve these equations numerically in the spectral representation \cite{my:zonal}. This model is equivalent to the CE2 \cite{ref:parker13, ref:parker14, ref:srinivasan12, ref:constantinou16}, but represents the dynamics of DWs in different (phase-space) variables. 

For comparison, we introduce the GO limit, which corresponds to $\smash{\text{max}\,(\lambda_{\text{DW}}/\lambda_{\text{ZF}}, \rho_s/\lambda_{\text{ZF}}) \ll 1}$. (Here, $\lambda_{\text{DW}}$ is the characteristic DW wavelength and $\lambda_{\text{ZF}}$ is the ZF spatial scale.) Then, \Eqs{eq:WM_W} and \eq{eq:WM_U} become
\begin{gather}
\pd_t W = \{\mc{H},W\}+2\Gamma W,\label{eq:WKE_W}
\\\textstyle
\pd_t U = \pd_y \int p_x p_y p_{D}^{-4} W\,d^2p/(2\pi)^2,\label{eq:WKE_U}
\end{gather}
which is called the iWKE model. Here, $W$ is understood as the phase-space distribution of driftons, $\mc{H}$ serves as their GO Hamiltonian, and $\Gamma$ serves as their damping rate. Specifically, 
\begin{gather}
\mc{H}= p_x U + p_x (U'' - \beta)/p_D^2,
\quad
\Gamma=-U'''p_xp_y/p_D^4.
\label{eq:WKE_HG}
\end{gather}
The tWKE has the same general form, \eqref{eq:WKE_W}, but with $\mc{H} = p_xU-\beta p_x/p_D^2$ and $\Gamma = 0$ \cite{ref:diamond05, ref:smolyakov99, ref:smolyakov00a, ref:smolyakov00b, ref:malkov01b, ref:kim02, ref:kaw02, ref:trines05, ref:miki12, ref:singh14}. (As mentioned previously, in contrast with the IWKE, the TWKE does not conserve the total energy and enstrophy of the DW-ZF system \cite{ref:parker16, my:zonal}.) We solve these equations numerically using discontinuous Galerkin methods implemented in the \textsc{gkeyll} code \cite{my:zonal, ref:shi17}.

\begin{figure}
\includegraphics[width=1\columnwidth]{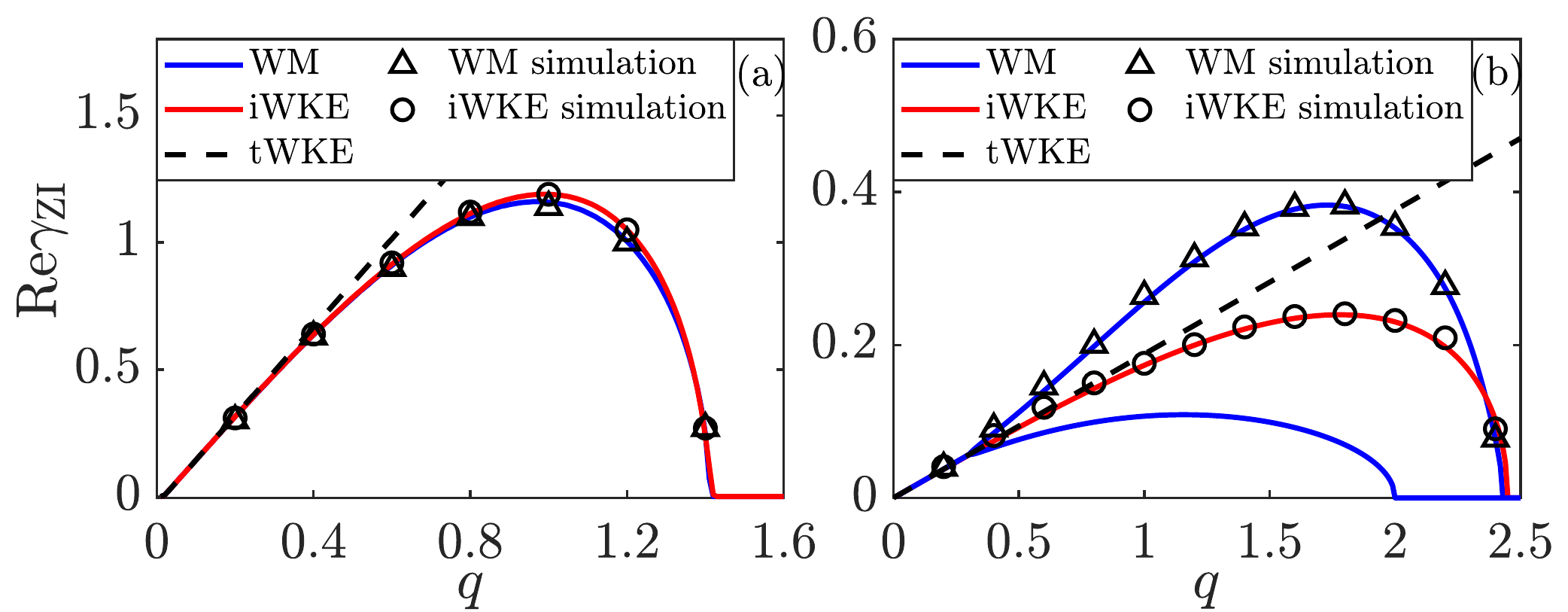}
\caption{$\gamma_{\rm ZI}(q)$ at $\beta=1$ for two equilibria: (a) $\mc{W}_1$ with $\mc{N} = 50$ and $p_f = 1$; (b) $\mc{W}_2$ with $k_x = 2$, $k_y = 1$, and $\mc{N}=100/(2\pi)^2$. Shown are the analytical results obtained from the WM (blue), iWKE (red), and tWKE (dashed) models, and the corresponding numerical results obtained from the WM (triangles) and iWKE (circles) simulations. The two blue lines in (b) correspond to two branches of $\text{Re}\,\gamma_{\rm TI}$. Only the fastest-growing mode is observed numerically.}
\label{fig:ZI}
\end{figure}

\section{Zonostrophic Instability}
\subsection{Linear ZI} 
First, we study the linear ZI, which is the formation of ZFs out of homogeneous DW turbulence with a given equilibrium Wigner function $\mc{W}(\vec{p})$. Within the WM approach, the ZI growth rate is found just like the kinetic dispersion of linear waves in a quantum plasma. Assuming $U = \text{Re}\,(U_q e^{iqy + \gamma_{\text{ZI}}t})$ and $\delta W = \text{Re}\,(W_q e^{iqy + \gamma_{\text{ZI}}t})$, one obtains \cite{my:zonal}%
\begin{multline}
\gamma_{\text{ZI}}=\int\frac{d^2p}{(2\pi)^2}\,\frac{qp_x^2p_y}{\gamma_{\text{ZI}}p_{D,+q}^2p_{D,-q}^2+2i\beta q p_x p_y}\\
\times\left[
\left(1-q^2/p_{D,-q}^2 \right)\mc{W}_{-q}-\left(1-q^2/p_{D,+q}^2\right)\mc{W}_{+q}\right],\label{eq:ZI_WM}
\end{multline}
where $\mc{W}_{\pm q}\doteq\mc{W}(p_x,p_y\pm q/2)$ and $p_{D,\pm q}^2\doteq 1 + p_x^2 + (p_y\pm q/2)^2$. For comparison, the iWKE predicts \cite{ref:parker16}
\begin{multline}\notag
1 = \int \frac{d^2p}{(2\pi)^2}\, \frac{q^2p_x^2p_{D}^{4}(1-4p_y^2/p_D^2)(1-q^2/p_D^2)}{\left(\gamma_{\text{ZI}}p_{D}^{4}+2i\beta qp_xp_y\right)^2}\,\mc{W}(\vec{p}).
\end{multline}
The tWKE result is obtained if one ignores $q^2/p_D^2$ in the second bracket in the numerator.

We considered two equilibria: $\mc{W}_1(\vec{p}) = 2\pi\mc{N} \delta(|\vec{p}| - p_{f})/p_{f}$ and $\mc{W}_2(\vec{p}) = \pi^2\mc{N} \sum_{m_{x,y} = \pm 1} \delta(p_x - m_x k_x)\delta(p_y - m_y k_y)$. Here, $\mc{N}[\mc{W}] = \int \mc{W}(\vec{p})\,d^2p/(2\pi)^2$ is the drifton density, or twice the DW enstrophy density \cite{my:zonal}, and $p_f$, $k_x$, and $k_y$ are constants. The simulations used $U(t = 0, y) = U_q \cos qy$ (with small $U_q$) and $W(t = 0, y, \vec{p}) = \mc{W}_{1,2}(\vec{p})$. The exponential growth of the perturbations in WM simulations agrees with \Eq{eq:ZI_WM} (\Fig{fig:ZI}). In contrast, the tWKE is adequate only at $q \ll 1$, and the corresponding $\gamma_{\rm ZI}$ has a maximum at the largest $q$ resolved numerically. Thus, the tWKE is inapplicable to modeling the ZI, as also noticed in \Refs{ref:parker16, my:zonal}. This means that the pioneering tWKE-based simulations of the ZI in \Ref{ref:trines05} were, at best, a qualitative demonstration of the effect. The iWKE is better, for it predicts that the ZI vanishes at $q \gtrsim 1$ and approximates $\text{Re}\,\gamma_{\rm ZI}$  reasonably well in the most important region (namely, $q \lesssim 1$) where $\gamma_{\rm ZI}$ has its maximum. But even so, the iWKE agreement with the full-wave theory is generally qualitative [\Fig{fig:ZI}(b)], and the WM model is more adequate.
 
\begin{figure}[t]
\includegraphics[width=1\columnwidth]{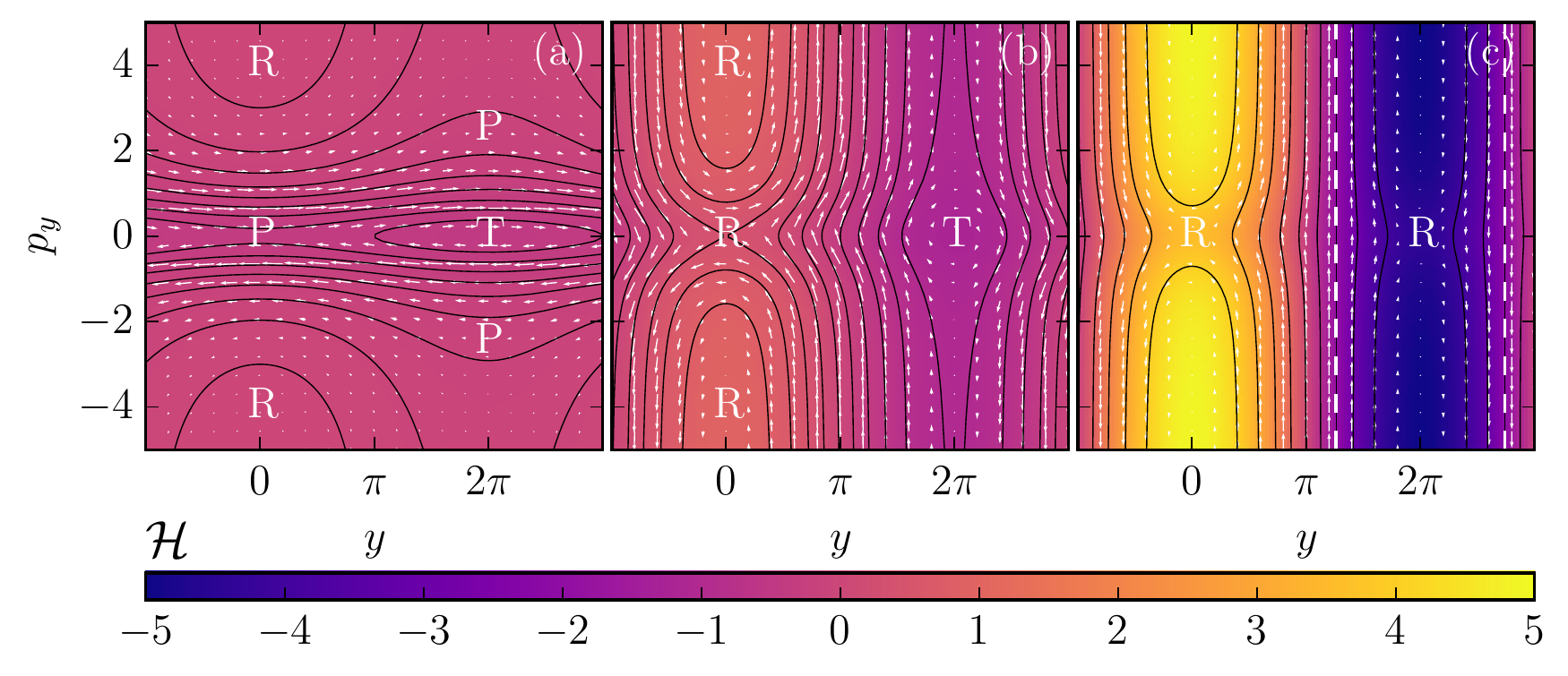}
\caption{Contour plots of $\mc{H}$ from the iWKE for $U = u_0\cos qy$ at $\beta=1$, $q=0.5$, and $p_x=0.5$: (a) Regime~1, $u_0 = 0.1$; (b) Regime~2, $u_0 = 2$; and (c) Regime~3, $u_0 = 10$. The arrows show the phase-space velocity given by \Eqs{eq:velocity} and \eq{eq:acceleration}. The labels P, T, and R denote passing, trapped, and runaway trajectories. The vertical dashed lines in (c) denote the locations where $U'' = \beta$.}
\label{fig:traj}
\end{figure}
 
\subsection{Nonlinear ZI} 
We also compare the GO and full-wave DW--ZF dynamics beyond the linear ZI. The former is elucidated by ray equations inferred from \Eqs{eq:WKE_W} and \eq{eq:WKE_HG}, 
\begin{gather}
\dot{y}
=\pd\mc{H}/\pd p_y
=2p_xp_y(\beta+q^2u_0\cos qy)/p_D^{4},
\label{eq:velocity}\\
\dot{p_y}
=-\pd\mc{H}/\pd y
=\left(1-q^2/p_D^2\right)p_x q u_0\sin qy,
\label{eq:acceleration}
\end{gather}
where we substituted a fixed ZF profile $U = u_0\cos qy$ for clarity. Three different topologies of the $(y, p_y)$ space are possible then, assuming $q < 1$ \cite{foot:details}. (At $q > 1$, the GO model is inapplicable, so it is not considered.) Regime~1 corresponds to weak ZFs, $u_0 < u_{c,1}\doteq \beta/(2-q^2)$ (\Fig{fig:traj}). This regime shows three types of trajectories: passing (labeled ``P''), trapped (labeled ``T''), and runaway (labeled ``R''), which extend to infinity along $p_y$ while being localized along $y$ \cite{foot:runaway}. Regime~2 corresponds to moderate ZFs, $u_{c,1} \leq u_0 < u_{c,2} \doteq \beta/q^2$. In this case, P-trajectories vanish but T- and R-trajectories persist. In Regime~3, only R-trajectories are left. This is the case of strong ZFs, $u_0 \geq u_{c,2}$. The latter is precisely the RK criterion \cite{ref:kuo49}, which has been known as a necessary condition of the ZF instability. Note that the RK parameter $\varrho \doteq u_0/u_{c,2}$ emerges in the iWKE but not in the tWKE, {where $u_{c,2}$ is infinite and hence Regime~3 is impossible.}

\begin{figure}[b]
\includegraphics[width=1\columnwidth]{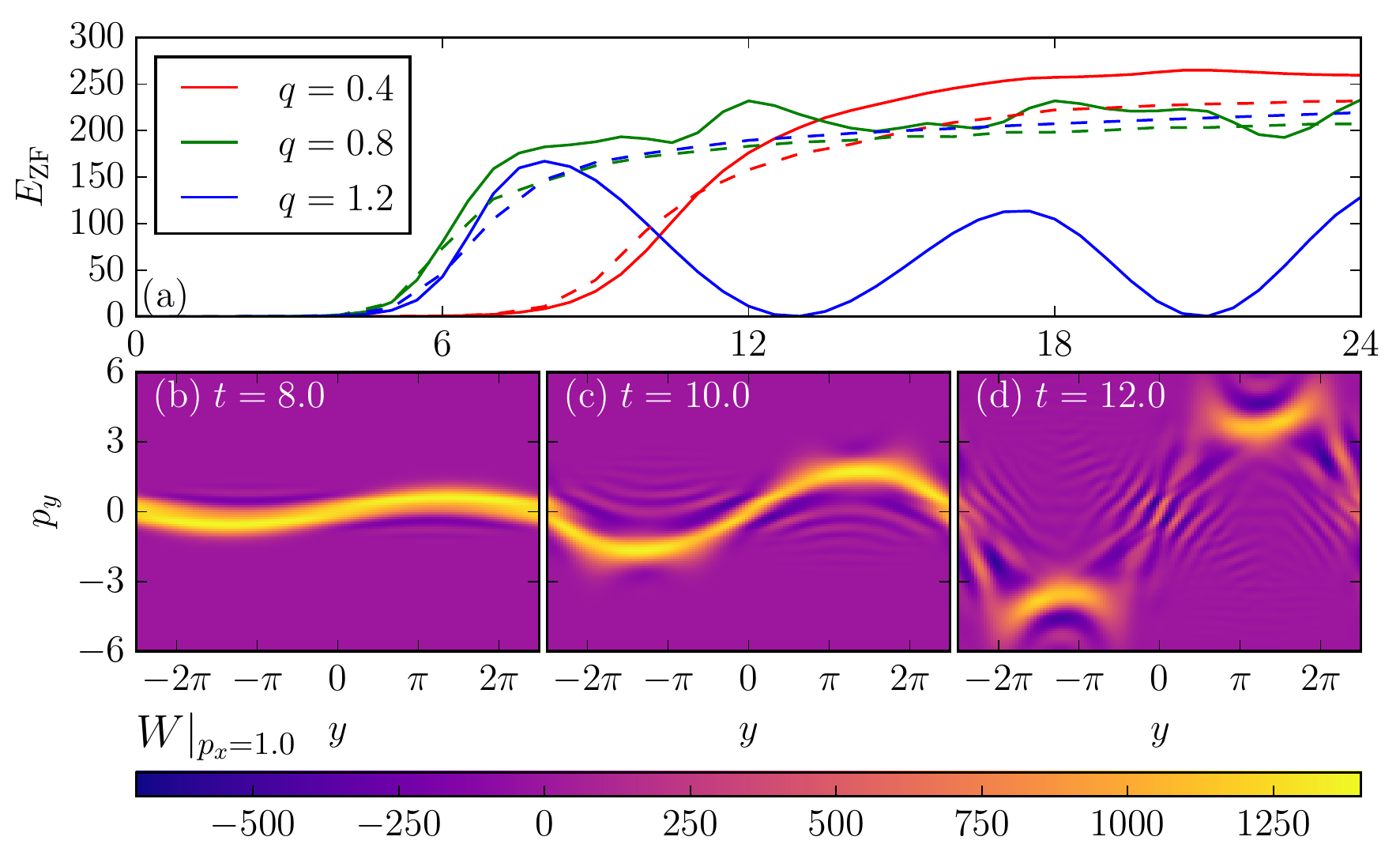}
\caption{Nonlinear simulations of the ZI with the same initialization as in \Fig{fig:ZI}(a). (a) The ZF energy $E_{\rm ZF} \doteq \int U^2\,dy/2$ versus $t$ for various $q$: iWKE model (dashed) and WM model (solid). At $q \lesssim 1$, the iWKE and WM models produce similar results. At $q \gtrsim 1$, WM simulations predict oscillations of $E_{\rm ZF}$. (b)-(d) Snapshots of $W$ from WM simulations ($q = 0.4$) for different $t$. The shape of the $\cap$ and $\cup$ structures is determined by the R-trajectories [Fig.~\ref{fig:traj}(c)]. Also see the movies in the supplementary material \cite{foot:sup}.}
\label{fig:nonlinzi}
\end{figure}
 
Since the total energy is conserved \cite{my:zonal}, the ZI eventually saturates. By taking moments of the iWKE, one also finds that $\pd_t U = [2(U'' - \beta)]^{-1}\,\pd_t \mc{N}$. Since the direction of phase-space flows is known (\Fig{fig:traj}), one can show from here \cite{foot:details} that, within the iWKE validity domain ($q < 1$), the profile of $U$ can \textit{only sharpen} with time. This implies that the ZI saturates monotonically, \ie never transfers its energy back to DWs. This is corroborated by both iWKE and WM simulations at $q \lesssim 1$; \ie the GO approximation is adequate in this case [\Fig{fig:nonlinzi}(a)]. In contrast, at $q \gtrsim 1$, full-wave effects are essential. In this regime, the iWKE and the tWKE are inapplicable, while WM simulations show that the ZI is eventually \textit{reversed}; \ie an intense ZF transfers its energy back to DWs (\Fig{fig:nonlinzi}). This results in predator--prey-type oscillations. They were also reported in the past \cite{ref:malkov01b, ref:miki12, ref:diamond94, ref:kim03} but were assumed to require ZF collisional damping. Our simulations show that this is not necessary. Besides, the oscillations were previously shown only within a tWKE-based model of drifton quasilinear diffusion, which assumes the GO limit and random small-amplitude ZFs. Neither of these assumptions holds in the regime when the oscillations occur in our simulations, so the WM approach is, in fact, necessary for accurate DW-kinetic modeling of these oscillations. Also, the importance of $q$ as a bifurcation parameter is consistent with our TI theory presented below.  

\section{Tertiary instability} 
Consider a system with initial conditions such that there is an intense ZF field and no DWs. Such ZF is subject to an instability of the Kelvin--Helmholtz type that we term TI. (The presence of DWs can affect the instability rate, as shown in \Refs{ref:parker14, ref:constantinou16} and in our discussion of the nonlinear ZI. We do not consider this effect here for it is hard to separate such TI from the nonlinear ZI.) This definition of the TI is different from that in \Refs{ref:rogers00, ref:rogers05}, where the TI was attributed to the ion-temperature gradient (absent in our model), but similar to those in the majority of relevant papers \cite{ref:kim02, ref:st-onge17, ref:singh16, ref:numata07}. In \Refs{ref:parker16, ref:numata07}, a connection was mentioned between the TI and the RK criterion, but the sufficient and necessary conditions for the TI were not explored analytically, and the mode structure has been unknown \cite{foot:kim}. Below, we propose two analytical and numerical calculations of the TI \cite{foot:details}. 

\begin{figure}[b]
\includegraphics[width=1\columnwidth]{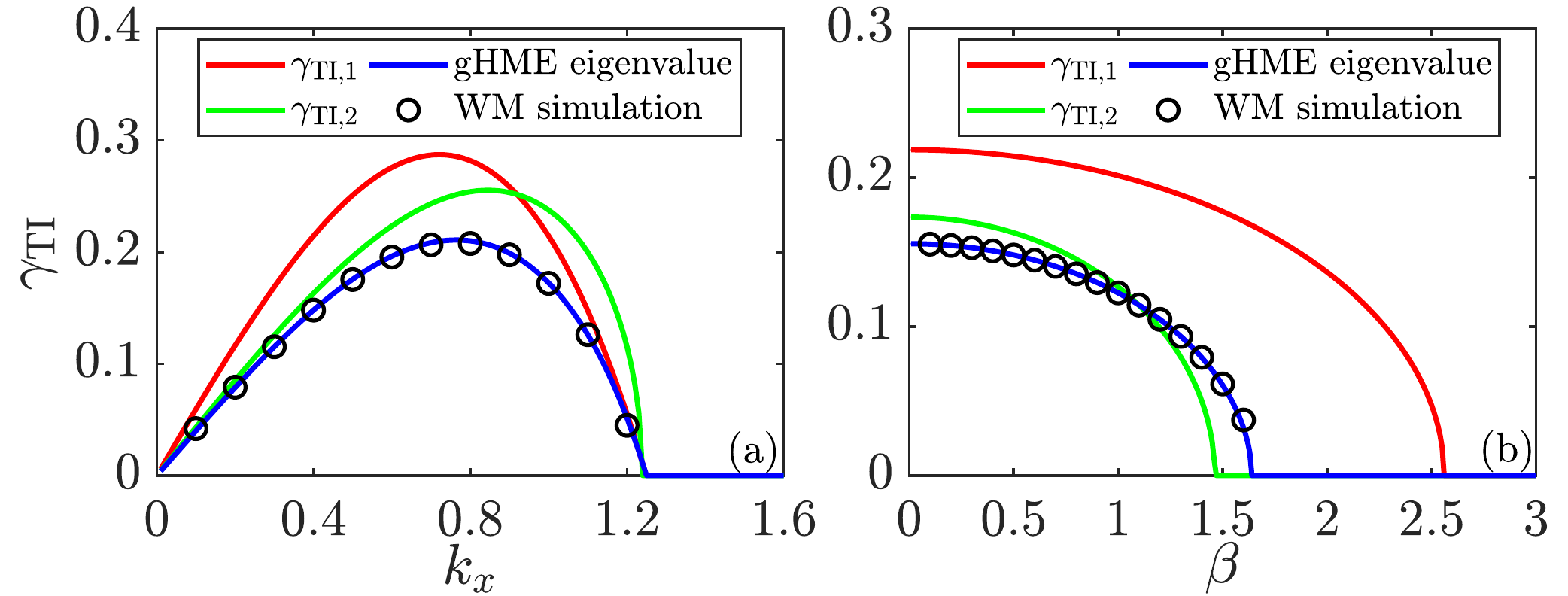}
\caption{(a) $\gamma_{\rm TI}(k_x)$ at ${\beta=0.5}$ and (b) $\gamma_{\rm TI}(\beta)$ at ${k_x=0.4}$. In both cases, $U(t = 0, y) = u_0\cos qy$, $u_0 = 1$, $q = 1.6$, and $\bar{q} = 0$. Shown are the analytical approximations \eq{eq:gamma_TI_1st} (red) and \eq{eq:gamma_TI} (green). Also shown are numerical solutions of the eigenvalue equation for $C$ (blue) and results of WM simulations (circles) with $\smash{W(t = 0, y, \vec{p})=W_1 \delta(p_x-k_x)e^{-p_y^2}}$ (with small $W_1$).}
\label{fig:Comparison}
\end{figure}

Let us consider $\widetilde{\varphi}= \text{Re}\,[\phi(y)e^{ik_x x - i\omega t}]$ and $C\doteq\omega/k_x$. Linearizing the gHME gives
\begin{gather}\label{eq:TI_linearized_gHME}
\left[d^2/dy^2-(1+k_x^2)-(U''-\beta)/(U-C)\right]\phi = 0.
\end{gather}
We assume $U = u_0 \cos qy$ and search for $\phi$ as a Floquet mode, $\phi = \psi(y)e^{i\bar{q}y}$, where $\psi(y + 2\pi/q) = \psi(y)$ and $\bar{q}$ is a constant restricted to the first Brillouin zone, $-q/2 \le \bar{q} < q/2$. Then, by following and correcting \cite{foot:details} Kuo's argument \cite{ref:kuo49}, we find that there are at most two unstable modes. The maximum of their growth rates, which we denote as the TI growth rate $\gamma_{\rm TI, 1} = \text{max}\,(k_x {\rm Im} C)$, is given by 
\begin{gather}\label{eq:gamma_TI_1st}
\gamma_{\rm TI,1} = |k_x u_0| \vartheta H(\vartheta) \sqrt{1-\varrho^{-2}},
\end{gather}
where $\vartheta \doteq 1-(\bar{q}^2 + 1 + k_x^2)/q^2$, $\varrho= u_0q^2/\beta$, and $H$ is the Heaviside step function. (The index 1 denotes that this is our first model of $\gamma_{\rm TI}$.) This growth rate is largest at $\bar{q} = 0$ and positive if $\varrho > 1$ and $q^2 > 1 + k_x^2 > 1$. Similar inequalities hold for nonsinusoidal ZF \cite{foot:details}. Hence, the necessary and sufficient conditions for the TI onset is twofold: (i) $\varrho \gtrsim 1$ and (ii) $q^2 \gtrsim 1$.  The latter implies a violation of the GO approximation. As a corollary, there is no TI in the GO limit. These findings also differ from those in Ref.~\cite{ref:kim02}, where the $\varrho$-dependence is missed.

\begin{figure}
\includegraphics[width=1\columnwidth]{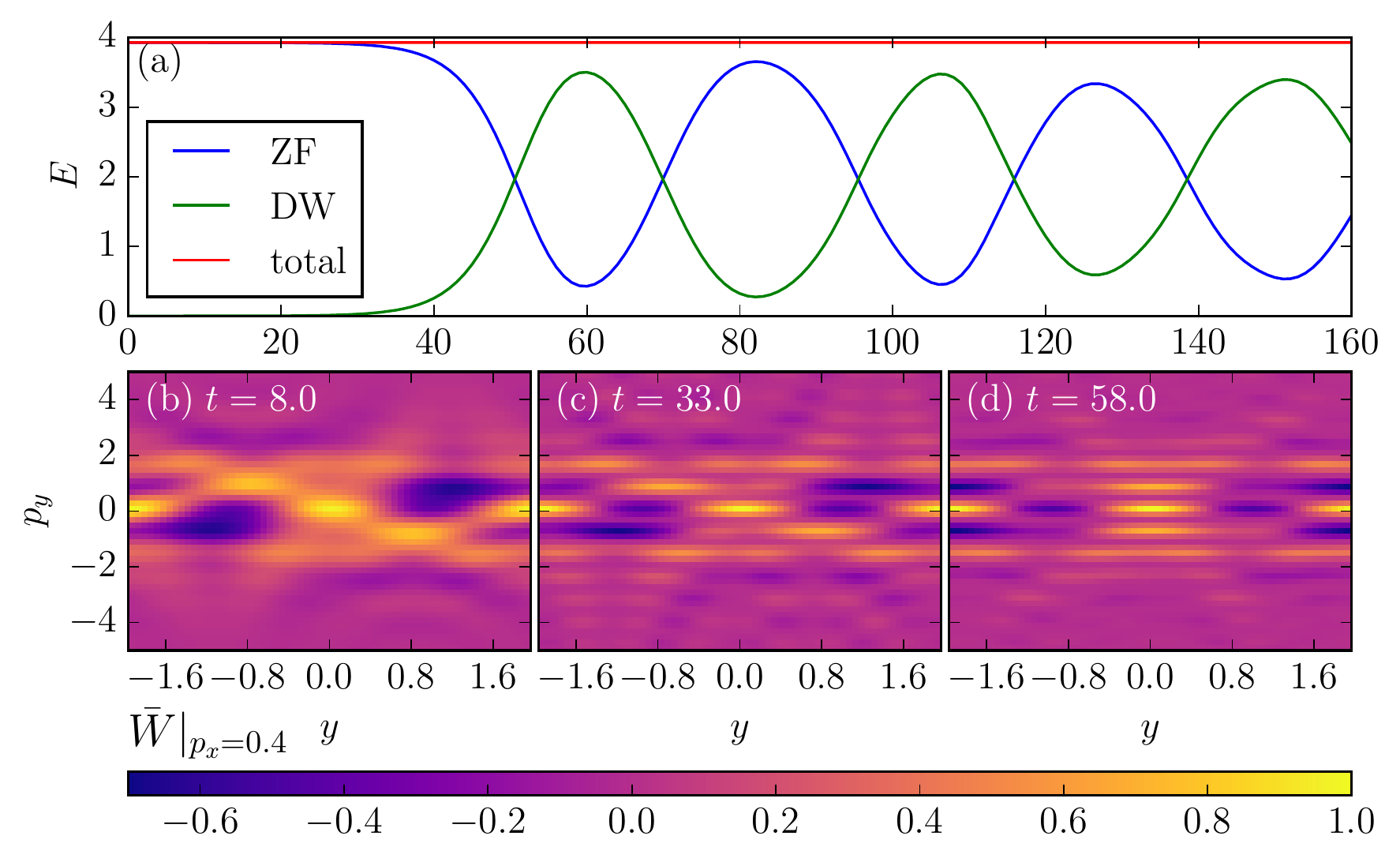}
\caption{Nonlinear simulations of the TI with the same initialization as in \Fig{fig:Comparison} ($\beta = 1$, $k_x = 0.4$). (a)~The energy of the ZF (blue), DWs (green) \cite{my:zonal}, and the total energy (red) versus $t$. (b)-(d) Snapshots of the normalized Wigner function $\bar{W}$ for different $t$. Figures (c) and (d) show the presence of multiple harmonics in the $p_y$ spectrum, which is because DWs are Floquet modes rather than point particles. Also, substantial regions of negative $\bar{W}$ are present. Hence, unlike in GO, $W$ cannot be understood as the probability distribution. This shows the importance of full-wave effects. The energy oscillations seen in figure (a) are correlated with the horizontal shifts of the phase space structures; compare figures (c) and (d). Also see the movie in the supplementary material~\cite{foot:sup}.}
\label{fig:ti}
\end{figure}

For comparison, we also calculated $\gamma_{\rm TI}$ numerically. First, we represent \Eq{eq:TI_linearized_gHME} as an eigenvalue problem, $\hat{A}^{-1}(U\hat{A}+\beta-U'')\psi = C\psi$, where $\hat{A}\doteq d^2/dy^2 + 2i\bar{q}d/dy - (\bar{q}^2 + 1 + k_x^2)$. Then, we adopted $\psi$ in the form $\psi = \smash{\sum_{m=-N}^{N}\psi_m e^{i mqy}}$, with truncation at a large enough $N$. Then, $C$ is found as an eigenvalue of a $(2N+1)\times(2N+1)$ matrix. As seen from \Fig{fig:Comparison}, \Eq{eq:gamma_TI_1st} is in reasonable agreement with the simulations but only when $\vartheta \ll 1$. In contrast, the WM approach allows for a calculation that extends to general $\vartheta$, namely, as follows. The numerical solution of the above eigenmode equation for $\psi$ can be used to calculate the eigenvector $\psi_m$, so we also obtain $\widetilde{w} = (\nabla^2-1) \widetilde{\varphi}$ and $W$. In the spectral representation $\mcc{W}(t, \lambda, \vec{p}) \doteq \int W(t, y, \vec{p})e^{-i\lambda y}\,dy$, the Floquet mode is a series of delta functions, $\mcc{W}(t, \lambda, \vec{p}) = \sum_{mn} \mcc{W}_{m,n}(p_x)\delta(\lambda - mq)\delta(p_y - nq/2)$, where $\mcc{W}_{m,n}$ decrease with $m$ and $n$. As an approximation, we retain only $\mcc{W}_{0,0}$, $\mcc{W}_{\pm 1,\pm 1}$, $\mcc{W}_{\pm 2, 0}$, and $\mcc{W}_{0, \pm 1}$. Then, from \Eq{eq:WM_W}, we obtain the eigenvalue
\begin{gather}\label{eq:gamma_TI}
\gamma_{\rm TI,2} = |k_x u_0| [\sqrt{2}(1 + \delta)]^{-1}\sqrt{1-\delta^2 - (2\delta^2\varrho^2)^{-1}},
\end{gather}
where $\delta \doteq (1 + k_x^2)/q^2$. The conditions for the TI onset within this model are $2\varrho^2 \delta^2 (1 - \delta^2) > 1$ and $q^2 > 1 + k_x^2$. This implies $\varrho > \sqrt{2}$ and $q^2 > 1$, which is in qualitative agreement with \Eq{eq:gamma_TI_1st}. Some discrepancy is explained by the fact that our series truncation is not a rigorous asymptotic approximation. For the same reason, $\gamma_{\rm TI,2}$ is not \textit{always} a better approximation of $\gamma_{\rm TI}$ compared to $\gamma_{\rm TI,1}$, but it does not require the smallness of $\vartheta$. Results of WM simulations of the TI are presented in \Fig{fig:ti}, which also illustrates the phase-space dynamics during the nonlinear stage. Our findings are in agreement with the direct numerical simulations reported in Ref.~\cite{ref:numata07}.

\section{Conclusions} 
We report full-wave phase-space modeling of the key basic effects associated with inhomogeneous DW turbulence and DW--ZF interactions. The turbulence is modeled as kinetics of an effective plasma where DWs act as quantumlike particles and the ZF velocity serves as their collective field. The drifton Hamiltonian is very different from that of conventional particles, so the phase-space dynamics is unusual and the applicability of the GO approximation is a subtle matter. Our findings show that traditional wave kinetics, which assumes the GO limit, misses essential physics in many aspects of the DW--ZF interaction problem. In contrast, the WM formulation is more robust and can be used as an efficient and intuitive tool for both analytical and numerical studies of DW turbulence. Our specific findings include a revised understanding of the nonlinear ZI and predator--prey oscillations and also a new theory of the TI within the gHME model. Applications of the WM formulation to other models of DW turbulence are anticipated in the future.

\begin{acknowledgments}
The authors thank J.~B. Parker and J.~A. Krommes for valuable discussions, E.~L. Shi for sharing \textsc{gkeyll} and an input script, and C.~F. Dong for assistance in using them. The work was supported by the U.S. DOE through Contract DE-AC02-09CH11466 and by Sandia National Laboratories. Sandia National Laboratories is a multimission laboratory managed and operated by National Technology and Engineering Solutions of Sandia, LLC., a wholly owned subsidiary of Honeywell International, Inc., for the U.S. Department of Energy's National Nuclear Security Administration under contract DE-NA-0003525. 
\end{acknowledgments}


\begin{thebibliography}{10}
\bibitem{foot:iwke}
The iWKE was originally derived using CE2 in \Ref{ref:parker16}, where it was termed CE2--GO, and rederived using the WM formalism in \Ref{my:zonal}, where it was termed WKE.
\bibitem{foot:details}
Details will be published in a separate paper.
\bibitem{foot:runaway}
For the lack of a better term, we call these driftons ``runaway'' by analogy
  with commonly known runaway electrons. However, the nature of the runaway
  effect for driftons is different as they are collisionless to begin with.
\bibitem{foot:kim}
Two calculations of $\gamma_{\rm TI}$ were proposed in \Ref{ref:kim02} but are
  incomplete. The first one, which uses a Floquet analysis, misses the
  stability threshold determined by $\varrho$. (The $q$-dependent threshold
  that we discuss in the main text is absent in \Ref{ref:kim02} because a
  different model of the electron response is assumed.) The second one, which
  uses the tWKE, is, in fact, a calculation of one of the ZI branches rather
  than of the TI. As we pointed out already, the tWKE cannot capture the TI in
  principle. In contrast, $\gamma_{\rm TI}$ reported in \Ref{ref:st-onge17} is
  consistent with our \Eq{eq:gamma_TI}.
\end{thebibliography}

\end{document}